\documentclass[letterpaper]{article}
\usepackage{amsmath}

\def\art#1{[\ref{#1}]}

\begin{document}

\title{\Large{On the ``The Kolmogorov-Smirnov test for the CMB'' by M.~Frommert, R.~Durrer and J.~Michaud}}
\author{
V.G.~Gurzadyan\footnote{Alikhanian National Laboratory, Yerevan, Armenia; gurzadyan@yerphi.am}
\footnote{Yerevan State University, Armenia.}
~and A.A.~Kocharyan\footnotemark[\value{footnote}]
\footnote{School of Mathematical Sciences, Monash University, Clayton, Australia;
\newline armen.kocharyan@monash.edu}
}

\maketitle

\begin{center}
ABSTRACT
\end{center}

In \art{FDM} the Kolmogorov-Smirnov (K-S) test and Kolmogorov stochasticity parameter (KSP) is applied to CMB data. Their interpretation of the KSP method, however, lacks essential elements. In addition, their main result on the Gaussianity of CMB was not a matter of debate in previous KSP-CMB studies which also included predictions on cold spots, point sources.   
\vspace{0.3in}

The study of CMB dataset in \art{FDM} is undertaken using the Kolmogorov-Smirnov test and the Kolmogorov stochasticity parameter. The former is a long known test, while the second one has been developed by Arnold in 2008-2009 \art{Arnold_KSP},\art{Arnold_UMN},\art{Arnold_MMS},\art{Arnold_FA} based on the work of Kolmogorov of 1933 \art{K}. Arnold not only revealed the informativity of KSP technique but also considered particular examples of number theory and dynamical systems. Based on Arnold's work, in \art{GK_KSP} KSP was applied to CMB where its power in separation of CMB from foreground Galactic disk signal has been shown, among other conclusions was the anomality of the Cold Spot and possibly of another region (Northern Cold Spot) \art{G2009}, prediction \art{G2010} of point sources (blazars) in CMB maps before they were identified by Fermi-LAT satellite. Recently KSP was shown to be able to detect galaxy clusters at X-ray observations \art{X}. 

K-S and KSP, irrespective to their similarities, do contain essential differences. Arnold defines KSP as an objectively measurable degree of randomness of observable events. Thus, it is much more than K-S test.

KSP is very similar to KS (Kolmogorov-Sinai) entropy $h(T)$ in Ergodic theory. If $h(T)>0$, then dynamical system $T$ is chaotic (we simplify the mixing/chaotic terminological link). However, we can use $h$ to compare different dynamical systems. We say $T_1$ is more chaotic than $T_2$ if $h(T_1)>h(T_2)>0$. $h(T)>0$ is the ``K-S test'' (chaotic or not), $h(T_1)>h(T_2)>0$ is the ``KSP'' (degree of chaoticity). 

We followed Arnold's approach and applied KSP to (highly) correlated data-sets CMB map \art{GK_KSP},\art{G2009}. We were not considering a temperature as 2D-map but as 1D scalars. 
$\delta T/T$ is a random (Gaussian) field. One can consider CMB map as 
\begin{itemize}
\item one sample from n-dimensional multivariate Gaussian distribution, or
\item $n$ samples from (different) univariate Gaussian distributions.
\end{itemize}

Of course, one may choose different variations of the mentioned two approaches e.g. by grouping some pixels into disks. 

Reducing KSP to KS takes the gloss off a very powerful and beautiful method. This is so obvious that we are surprised to see papers not fully appreciating their difference\footnote{The identity of K-S and KSP was assumed in \art{SKN} with a claim as if KSP can be applied only to non-correlated data (as K-S test); the invalidity of such assumptions was pointed out in \art{K1}, see also \art{K2}.}.

The main result in \art{FDM} is the Gaussianity of CMB.  The same is concluded at KSP-CMB studies in \art{GK_KSP},\art{G2009}, e.g. it is clearly stated (p.345 in \art{G2009}) that the Galactic disk with Kolmogorov's $\Phi=1$ implies the non-Gaussianity of its radiation, i.e. KSP separates the Gaussian CMB from non-Gaussian foreground disk.

Estimations of $\Phi$ for particular areas reflect the correlations on the scale of the region. 
The estimation \art{wrandom} of 20\% random fraction of the CMB signal by no means contradicts its Gaussian nature. The concise style of \art{GK_KSP},\art{wrandom} cannot be enough reason for attibuting them statements not existing there.
The K-S test has been already applied to CMB in \art{Cruz}.

\end{document}